\documentclass[conference]{IEEEtran}
\IEEEoverridecommandlockouts

\usepackage{cite}
\usepackage{amsmath,amssymb,amsfonts}
\usepackage{algorithmic}
\usepackage{graphicx}
\usepackage{textcomp}
\usepackage{xcolor}
\usepackage{url}
\usepackage{newtxtext,newtxmath}
\def\BibTeX{{\rm B\kern-.05em{\sc i\kern-.025em b}\kern-.08em
    T\kern-.1667em\lower.7ex\hbox{E}\kern-.125emX}}

\begin{document}

\title{Mini-SFC: A Comprehensive Simulation Framework for Orchestration and Management of Service Function Chains}

\author{\IEEEauthorblockN{Xi Wang, Shuo Shi, Chenyu Wu}
\IEEEauthorblockA{\textit{School of Electronics and Information Engineering} \\
\textit{Harbin Institute of Technology}\\
Harbin, China \\
wangxi\_chn@foxmail.com, crcss@hit.edu.cn, wuchenyu@hit.edu.cn}
}

\maketitle

\begin{abstract}
In the continuously evolving cloud computing and network environment, service function chain (SFC) plays a crucial role in implementing complex services in the network with its flexible deployment capabilities. To address the limitations of existing SFC simulation tools, this paper introduces Mini-SFC, a modular simulation framework that supports both numerical and container-based virtual simulations, while also supporting online dynamic topology adjustments. As an open-source platform emphasizing user-friendliness, Mini-SFC facilitates rapid algorithm verification and realistic service deployment validation. By simplifying module design and providing standardized solver interfaces, Mini-SFC significantly shortens the learning curve for researchers and enhances the flexibility and scalability required for advanced SFC management and optimization. For readers interested in exploring or utilizing Mini-SFC, more information is available on the official project page.
\end{abstract}


\section{Introduction}

In the context of the intertwined development of cloud computing and networking, service function chain (SFC) offers a method to implement complex business logic through the sequential connection of network services. It enables network administrators to efficiently configure and manage various network functions, such as firewalls, load balancers, and intrusion detection systems, thereby enhancing network resource utilization and service quality\cite{SFC}. Particularly, in cloud environments, SFC is often used in conjunction with Network Functions Virtualization (NFV), which replaces traditional dedicated hardware with Virtual Network Functions (VNFs) on general-purpose servers for dynamic deployment and adjustment, significantly increasing the flexibility and scalability of cloud services. Meanwhile, the introduction of Software-Defined Networking (SDN), which separates the control plane from the data plane, provides capabilities for centralized management and dynamic traffic control, laying the foundation for network automation and intelligence in cloud computing environments\cite{SDN}.

Through network virtualization technology, SFC significantly improves the flexibility and scalability of the network, enabling dynamic adaptation to ever-changing demands. With the advancement of communication technology, more and more research focuses on deploying SFC in complex scenarios, such as the space-air-ground integrated network (SAGIN) to support 6G networks\cite{SAGIN}. However, the current SFC simulation tools face limitations such as difficulty in use and lack of support for dynamic topology\cite{Mini-NFV}. To provide a better simulation experimental platform, Mini-SFC introduces a modular simulation framework aimed at overcoming the limitations of existing tools. It provides numerical and container-based virtual simulation, supporting rapid algorithm verification and real service deployment, accurately simulating interactions and resource competition. Mini-SFC can be deployed on a single machine, making it possible to complete experiments with a low-cost environment, which is very helpful for research teams with limited resources. The Mini-SFC is an open-source project. The documentation and examples can be found at\cite{Mini-SFC}. 

Notably, this work differs fundamentally from our prior publication\cite{PSO_SFC}. The earlier study focused on designing and optimizing a specific SFC embedding and migration algorithm for UAV IoT scenarios, with core contributions in algorithmic innovation and performance improvement. In contrast, the current paper centers on the design and implementation of a general-purpose simulation framework that supports SFC research across diverse scenarios. Mini-SFC is not tied to any specific algorithm; instead, it provides a flexible, low-cost, and user-friendly platform to verify, compare, and optimize various SFC-related algorithms. This shift from \textit{algorithm research} to \textit{tool development} addresses the critical lack of versatile simulation tools in the field, which is the primary innovative contribution of this work. The open-source nature of Mini-SFC aims to promote transparency and collaboration. It is hosted on GitHub\cite{Mini-SFC} and provides a tutorial wiki for users to use and contribute new solvers, as well as positive responses from developers to technical queries. This ongoing maintenance model guarantees that researchers receive timely support and updates, enhancing the framework's reliability for long-term research.

\section{Related Works}

In the field of SFC management research, different experimental platforms possess their own characteristics and application scenarios. Numerical simulation, container-based virtual simulation, and real deployment are three primary forms of such platforms. Numerical simulation quickly validates theoretical hypotheses through mathematical models and algorithms, suitable for large-scale scenario testing; container-based virtual simulation assesses real-world operational performance by constructing realistic virtual environments; whereas real deployment runs services directly in a production environment to evaluate system performance under the most authentic conditions but faces challenges like high costs, complex deployment, and low flexibility.

Virne\cite{Virne} is a numerical simulation tool supporting SFC deployment, integrating numerous cutting-edge deployment algorithms and is user-friendly, yet it does not support simulation or dynamic topology adjustment. Apollo\cite{Apollo} is an SFC orchestrator Microsoft uses to operate directly in its server cluster management, but lacks specific implementation details. MD-VNs\cite{MD-VNs} can simultaneously perform numerical and OpenStack-based simulations but neither supports dynamic topologies nor is open-source, limiting its scope of application. PyCloudSim\cite{PyCloudSim} simulates SFCs by introducing multi-core CPU, network interface, and priority process models, but it lacks support for numerical simulations and online dynamic topologies at large scale. Both Mini-nfv\cite{Mini-NFV} and vim-emu\cite{vim-emu} are functional extensions of Mininet using Open vSwitch to build virtual networks. Mini-nfv simulates VNFs constituting SFCs on KVM-based hosts with a focus on implementing VNFFG (VNF Forwarding Graph) models based on TOSCA, though it struggles with effectively managing and constraining resources of individual VNFs. Vim-emu uses Docker to simulate VNFs comprising SFCs and integrates with the Open Source MANO (OSM) platform for SFC orchestration functionality. However, due to the complex of the OSM project, researchers merely needing to validate their designed orchestration algorithms face extremely high learning curves.

\begin{table}[htbp]
	\caption{Comparison of SFC Simulation Tools}
	\begin{center}
		\begin{tabular}{|p{1.3cm}|p{1.2cm}|p{1.2cm}|p{1cm}|p{0.8cm}|p{0.8cm}|}
			\hline
			\textbf{Project}&\textbf{Numerical}&\textbf{Virtual}&\textbf{Dynamic}&\textbf{Open}&\textbf{Easy}\\
			\textbf{Name}&\textbf{Simulation}&\textbf{Simulation}&\textbf{Topology}&\textbf{Source}&\textbf{to Use}\\
			\hline
			Virne		&Yes	&No		&No		&Yes	&Yes\\
			Apollo		&No		&Yes	&No		&No		&No\\
			MD-VNs		&Yes	&Yes	&No		&No		&No\\
			PyCloudSim	&No		&Yes	&No		&Yes	&Yes\\
			Mini-nfv	&No		&Yes	&No		&Yes	&Yes\\
			vim-emu		&No		&Yes	&Yes	&Yes	&No\\
			Mini-SFC	&Yes	&Yes	&Yes	&Yes	&Yes\\
			\hline
		\end{tabular}
		\label{tab1}
	\end{center}
\end{table}

As shown in Table \ref{tab1}, the Mini-SFC simulation framework offers a distinctive approach compared to existing tools. Mini-SFC supports both numerical simulations and container-based virtual simulations, along with dynamic topology adjustments. Designed as an open-source platform with user-friendliness in mind, it enables researchers to conduct large-scale scenario tests efficiently, aiding in the swift validation of algorithm performance and optimization strategies. The framework's container-based virtual simulation capability allows for service deployment in environments that closely mimic real-world conditions, effectively simulating complex interactions and resource contention issues.

\section{Mini-SFC Architecture}

\subsection{Core Design Philosophy}

\begin{figure}[htbp]
	\centerline{\includegraphics[width=0.5\textwidth]{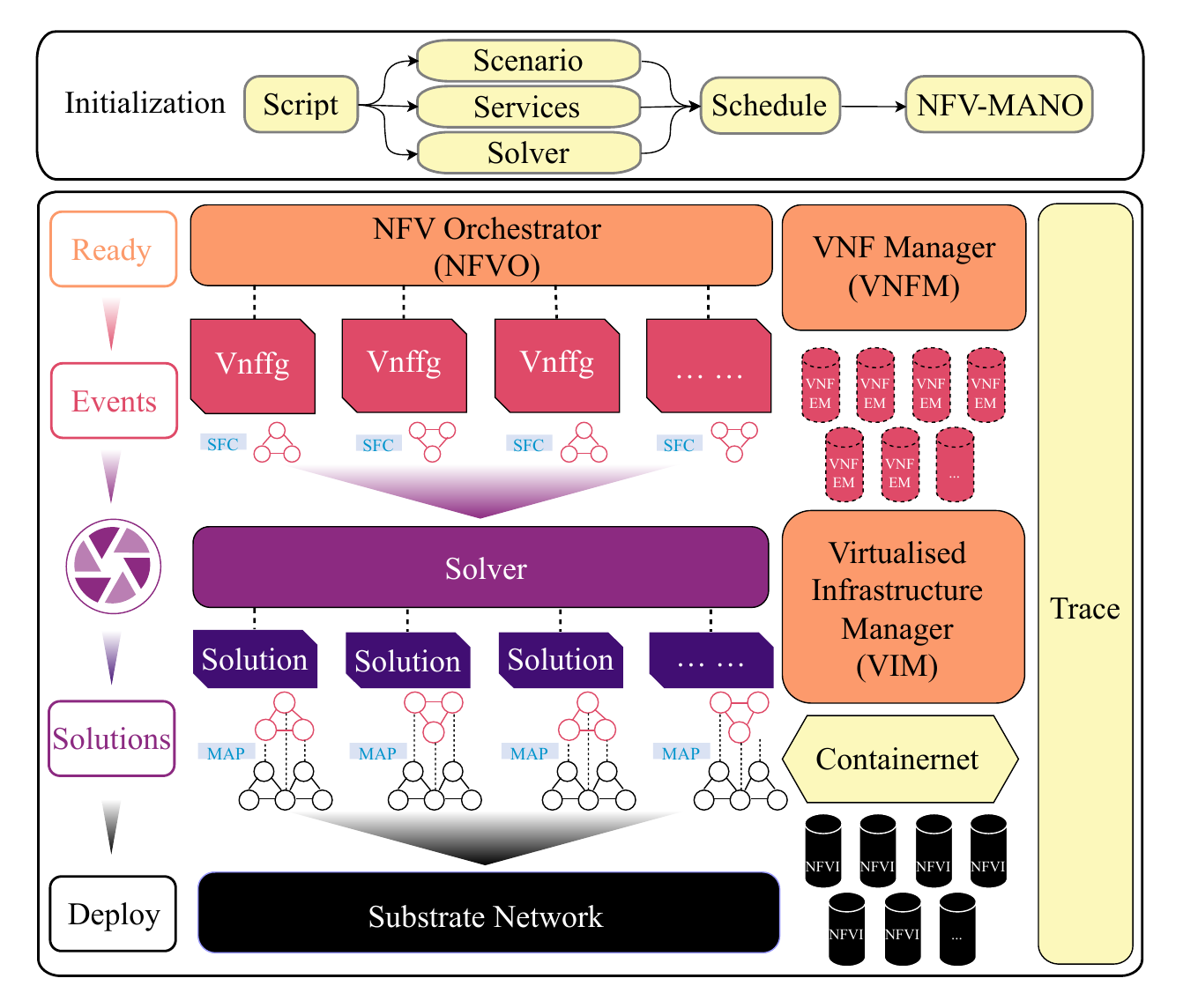}}
	\caption{The Mini-SFC architectural diagram.}
	\label{fig_1}
\end{figure}

The Mini-SFC adopts a design philosophy centered on flexibility, scalability, and simulation accuracy. This platform uniquely integrates numerical and container-based simulations while emphasizing modular design to facilitate rapid iteration and customization. The architecture of Mini-SFC, as depicted in Fig. \ref{fig_1}, adopts a layered structure designed to streamline component interactions. This design not only enhances code reusability and maintainability but also significantly boosts user flexibility.

\subsubsection{Numerical and Container-Based Simulation Integration}

Mini-SFC combines numerical and container-based simulation through a user-selectable control variable. By default, SimPy is used for discrete event simulation, focusing on SFC deployment algorithms and their validation in large-scale virtual networks. In this mode, topology information is maintained using Networkx, with infrastructure resources abstracted as node attributes and link resources as edge attributes. VNFs' resource occupation and release modify these topological data, akin to methods seen in Virne and Matlab-based tools, offering easy scalability and verification.

Alternatively, selecting container-based simulation activates real-time event progression via Containernet, utilizing Open vSwitch and Docker containers. This mode simulates actual service times defined for SFCs, creating and destroying Docker containers to represent VNF resource dynamics. Service operations, including data transmission between VNFs, are executed in real-time, providing a more accurate reflection of operational environments than PyCloudSim\cite{PyCloudSim}, which uses simulated processes and packets. While Mini-SFC is designed for flexibility and low-cost deployment on a single machine , it is important to note potential performance overhead and scalability constraints, particularly in container-based mode. In container-based simulation, each VNF and user equipment (UE) is instantiated as a separate Docker container, introducing non-negligible overhead in CPU scheduling, memory allocation, and inter-container network communication. On a typical single machine with 8-core CPU and 32GB RAM, experiments show that concurrent operation of more than 30 containers may lead to noticeable latency in VNF startup and increased jitter in data transmission between VNFs, primarily due to resource contention.

This dual-mode approach ensures Mini-SFC supports both theoretical validations at early design stages and practical operation evaluations during final integration testing, making it a useful tool for SFC orchestration and management research from concept to deployment.

\subsubsection{Reference to MANO Architecture}

Mini-SFC incorporates the core concepts from the Management and Orchestration (MANO) architecture, significantly simplifying its module design while preserving solver interfaces for executing SFC deployment algorithms. By adopting MANO's standardized definitions of each module, Mini-SFC enhances system modularity and scalability. In the MANO, the NFVO manages services and network slicing, VNFM oversees the lifecycle management of VNFs, and VIM controls both physical and virtual resources. This standardized approach not only increases system flexibility but also facilitates smoother collaboration between different modules.

However, deploying OSM directly like vim-emu\cite{vim-emu} can introduce significant complexity due to their extensive configuration requirements and the need for a thorough understanding of SFC deployment algorithms and their execution interfaces. For researchers looking to test algorithm effectiveness, these complexities often pose a substantial barrier. To address this, Mini-SFC simplifies module design and provides unified interfaces with standardized solver input and output formats. Researchers need only provide their algorithm inputs and outputs in a standard format to seamlessly integrate into the Mini-SFC framework and begin execution. This streamlined approach greatly reduces the learning curve, enabling researchers to quickly engage in actual research work without being bogged down by platform's complexities.

\subsection{Detailed Explanation of Modular Design}

\subsubsection{MANO Modules}

In Mini-SFC, MANO modules follow the design principles of their architecture, facilitating efficient resource management and service orchestration through simplified interfaces. These modules aim to streamline the collaboration between different components of the system.

\begin{itemize}
	\item NFVO (NFV Orchestrator): Manages the lifecycle and deployment of SFCs to meet diverse service needs.
	\item VNFM (VNF Manager): Manages the lifecycle and deployment of VNFs, ensuring their correct operation.
	\item VIM (Virtualized Infrastructure Manager): Oversees the management of physical and virtual resources on each NFVI (NFV Infrastructure).
	\item UEM (User Equipment Manager): Handles data plane operations, focusing on user data packet forwarding within virtual networks.
\end{itemize}

This modular approach ensures smooth interaction between components, providing a flexible platform for researchers to test and validate algorithms in a simplified environment.

\subsubsection{Event Module}

Mini-SFC uses an event-driven mechanism to advance the simulation process. This design allows the system to simulate dynamic changes that occur in the real network operating environment, triggering specific operational procedures through key events, making the simulation more flexible.

\subsubsection{Net Module}

The net module is the core module in Mini-SFC that manages and schedules various components, providing the external interfaces for the simulator's initialization, start, and end.

\subsubsection{Topo Module}

The topo module is inherited from Networkx and is mainly used to store the data required by the simulator in the form of graphs, such as the substrate network and its resource capacity, the service network and its resource request volume, etc.

\subsubsection{Trace Module}

The trace module facilitates performance evaluation and problem diagnosis by collecting critical data during simulation. The ability to customize trace modules to monitor state values of interest while running is critical for monitoring algorithm performance and resolving problems quickly.

\subsubsection{Solver Module}

The solver module provides a framework for developing custom algorithms, supporting the implementation and testing of different optimization strategies for SFC scheduling. Rooted in Mini-SFC's discrete event-driven workflow, this module simplifies algorithm integration through a standardized input/output mechanism aligned with the platform's event-triggered operation. When an SFC deployment or migration event is triggered (e.g., a new service request or topology change), the solver automatically receives two key inputs: the details of the current SFC event (such as VNF types, resource demands, and QoS constraints) and the real-time network state (including available node/link resources and current topology) from the substrate network. For output, the solver returns a structured mapping table template that specifies VNF-to-node assignments, inter-VNF routing paths, and resource allocation details. This template is directly parsable by MANO components (NFVO and VNFM), which then automatically execute the deployment or migration strategy without additional configuration.

Users only need to comply with the input/output specifications of the framework and can implement their own SFC deployment and migration algorithms by inheriting the module. This design eliminates the need for researchers to handle low-level event scheduling or MANO integration logic, drastically reducing the experimental cost of validating new algorithms compared to platforms with complex interfaces (e.g., vim-emu's reliance on OSM workflows\cite{vim-emu}), thereby embodying Mini-SFC's emphasis on user-friendliness and flexibility.

Through this modular design, Mini-SFC achieves high flexibility and scalability, embodying a thoughtful approach to support research in orchestration and management of SFC. The system is designed to be easy to use, catering to the evolving needs of researchers.

\section{Mini-SFC Usage Example}

\begin{figure}[htbp]
	\centerline{\includegraphics[width=0.5\textwidth]{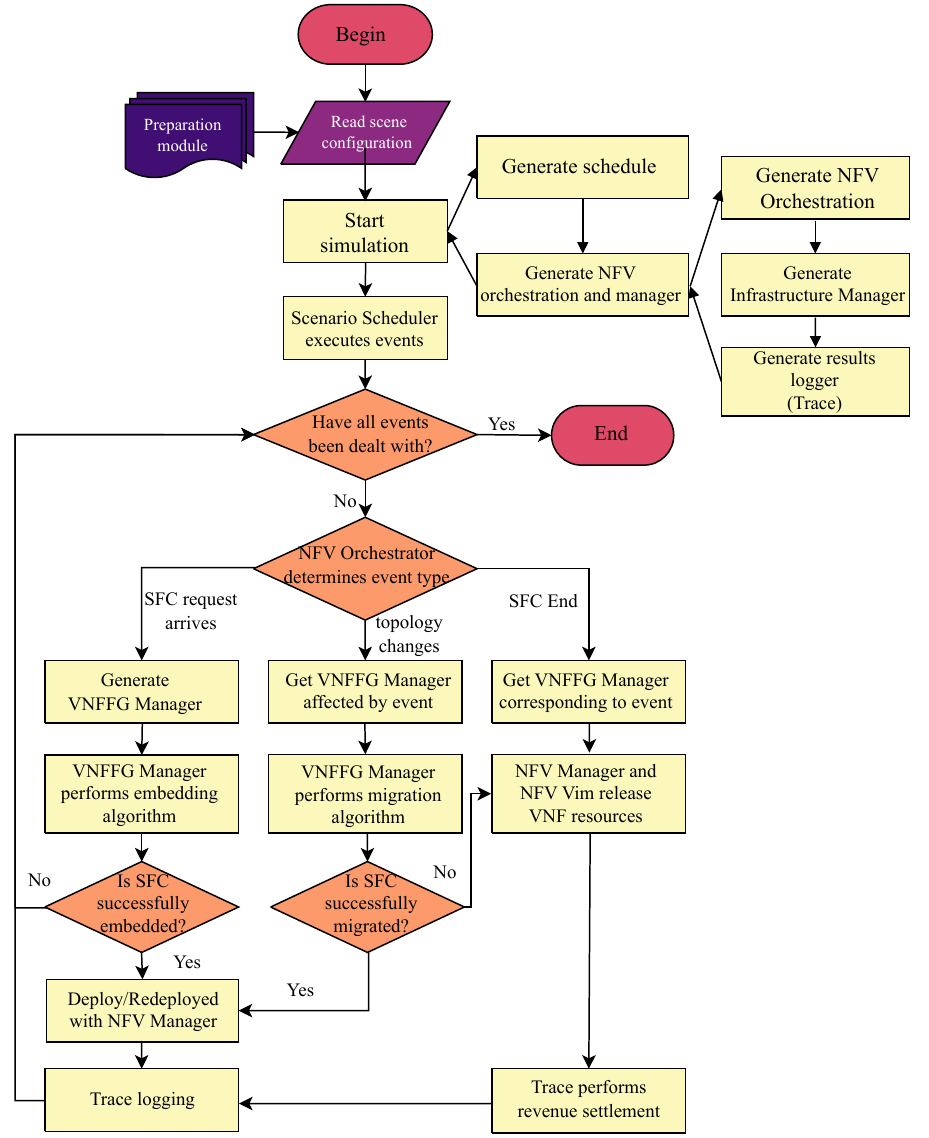}}
	\caption{The execution flow of Mini-SFC.}
	\label{fig_2}
\end{figure}

Mini-SFC offers a flexible and powerful simulation platform to validate and optimize the orchestration and management strategies of SFCs. Drawing inspiration from the script-driven - simulation engine - log analysis workflow seen in tools like Mininet and NS3. The execution flow of the entire framework is illustrated in Fig. \ref{fig_2}.

We will demonstrate the advantages of Mini-SFC through two simple use examples, highlighting its support for numerical and virtual simulation, and its ease of use with custom orchestration algorithms.

\subsection{Numerical Simulation Example}

Performing numerical simulations using Mini-SFC generally involves the following steps:

\subsubsection{Definition of the Substrate Network Topology}

Define parameters such as time list, adjacency matrix dictionary, weight matrix dictionary, node resource dictionary, and link resource dictionary, and set the initialization state of the substrate network in the simulation scenario. These dictionaries are indexed by time and are used to represent the network state at different times, thus achieving a dynamic scene setting.

\begin{tabbing}
	\hspace*{20pt}\= \kill
	\textbf{from} minisfc.topo \textbf{import} SubstrateTopo \\
	substrateTopo = SubstrateTopo(timeList, \\
	\> adjacencyMatDict, weightMatDict, \\
	\> nodeResourceDict, linkResourceDict)
\end{tabbing}

\subsubsection{Definition of the Service Function Chains}

Define parameters including a list of SFCs' ID, lifecycle dictionary, endpoint dictionary, VNFs request dictionary, and Quality of Service (QoS) requirements dictionary. These dictionaries are indexed by the SFC's ID to represent the service needs of different users.

\begin{tabbing}
	\hspace*{20pt}\= \kill
	\textbf{from} minisfc.topo \textbf{import} ServiceTopo \\
	serviceTopo = ServiceTopo(sfcIdList, \\
	\> sfcLifeTimeDict, endPointDict, \\
	\> vnfRequstDict, qosRequesDict)
\end{tabbing}

\subsubsection{Configuration of the Virtual Network Function Manager}

Add the required VNF instances as templates to the library and set the virtual bandwidth between them.

\begin{tabbing}
	\hspace*{20pt}\= \kill
	\textbf{from} minisfc.mano.vnfm \textbf{import} VnfManager, VnfEm \\
	nfvManager = VnfManager() \\
	vnfEm\_template = VnfEm(**\{'vnf\_id':0,'vnf\_cpu':0.2,\\
	\> 'vnf\_ram':64\}) \\
	nfvManager.add\_vnf\_into\_pool(vnfEm\_template)\\
	nfvManager.add\_vnf\_service\_into\_pool(0,1,**{"band":20})
\end{tabbing}

\subsubsection{Selection of Solver and Launching the Simulation}

Choose a solver to execute the SFC deployment algorithm (e.g., RandomSolver or a user-defined algorithm), create and initiate the Mini-SFC network simulation by integrating all components into the engine.

\begin{tabbing}
	\hspace*{20pt}\= \kill
	\textbf{from} minisfc.solver \textbf{import} RandomSolver \\
	\textbf{from} minisfc.net \textbf{import} Minisfc \\
	sfcSolver = RandomSolver(substrateTopo,serviceTopo) \\
	net = Minisfc(substrateTopo,serviceTopo,\\
	\> nfvManager,sfcSolver) \\
	net.start()\\
	net.stop()
\end{tabbing}

In this example, we configured the substrate network with three nodes, structured in a chain topology. These nodes respectively have core counts of [2, 4, 2], memory sizes of [256MB, 512MB, 256MB], and link bandwidths of 100 Mbps. During the network operation, two SFCs were requested by users, with lifecycles spanning from 5 to 25 seconds and 10 to 50 seconds after the start of the simulation, respectively. Each SFC requests for 3 VNFs selected from the VNF Manager. The simulation stops once all service requests are processed.

\begin{figure}[htbp]
	\centerline{\includegraphics[width=0.5\textwidth]{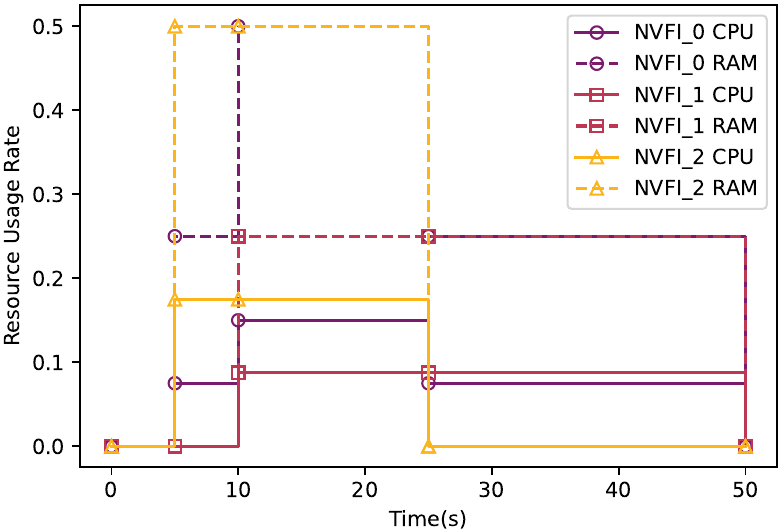}}
	\caption{The resource load status of NVFIs in the numerical simulation example.}
	\label{fig_3}
\end{figure}

During the simulation execution, the resource load status of the infrastructure manager on the substrate network is shown in Figure \ref{fig_3}. It can be observed that the MANO deploys more user-requested VNFs on node NVFI\_2, leading to its CPU and RAM utilization rate being the highest among all nodes. Meanwhile, it is also evident that at the 25th and 50th second, as services conclude, the network resources occupied by the two SFCs are sequentially released, eventually bringing the network back to its initial state.

Notably, Mini-SFC facilitates systematic exploration of hyperparameter impacts, which is critical for optimizing settings in specific use cases. For instance, in numerical simulation mode, key hyperparameters such as node CPU/RAM capacities, link bandwidth limits, and SFC lifecycle thresholds can be adjusted to observe their effects on deployment performance. By adopting a control-variable approach—e.g., incrementally increasing node CPU capacity while fixing other parameters—researchers can quantify how resource constraints influence SFC embedding success rates and resource utilization (as shown in Figure \ref{fig_3}). This flexibility allows targeted optimization for scenarios like resource-constrained edge networks or large-scale cloud environments, which is less straightforward in tools like Virne\cite{Virne} due to their fixed topology and resource model limitations.

This numerical simulation example demonstrates Mini-SFC's efficiency in theoretical verification scenarios. By defining substrate network topology, SFC requirements, and VNF resource parameters, researchers can quickly validate algorithms in large-scale scenarios, leveraging its SimPy-based discrete event simulation and Networkx-driven topology management. Compared to existing tools, Virne\cite{Virne} supports numerical simulation but lacks dynamic topology adjustment, making it unsuitable for time-evolving network scenarios. MD-VNs\cite{MD-VNs} integrates numerical and OpenStack-based simulations but is neither open-source nor dynamic-topology-aware, limiting extensibility. PyCloudSim\cite{PyCloudSim} omits numerical simulation entirely. Mini-SFC’s uniqueness lies in combining numerical simulation with dynamic topology support while maintaining open-source accessibility and a low learning curve, enabling researchers to validate algorithms across static-to-dynamic scenarios within a single tool—an capability unmet by existing alternatives.

\subsection{Container-based Simulation Example}

In this example, the steps for \emph{Definition of the Substrate Network Topology} and \emph{Definition of the Service Function Chains} are identical to those in Example A and thus will not be repeated here.

\subsubsection{Configuration of the Virtual Network Function Manager}

Define the VNFs and add them to the manager pool. Unlike in Example A, this configuration requires providing a command template for launching VNF containers. Additionally, specific container images must be supplied to ensure that the corresponding VNF Docker containers can start correctly. Here we use a mirror prepared in advance. A FastAPI-based web server was deployed on the mirror, performing multiple inversions of the incoming matrix to simulate high computational load requirements.

\begin{tabbing}
	\hspace*{20pt}\= \kill
	\textbf{from} minisfc.mano.vnfm \textbf{import} VnfManager,VnfEm \\
	nfvManager = VnfManager() \\
	template\_cmd = "python run\_command.py \\
	\> --vnf\_name = \$vnf\_name \\
	\> --vnf\_type = \$vnf\_type\\
	\> --vnf\_ip = \$vnf\_ip --vnf\_port = \$vnf\_port \\
	\> --vnf\_cpu = \$vnf\_cpu --vnf\_ram = \$vnf\_ram \\
	\> --vnf\_rom = \$vnf\_rom" \\
	vnfEm\_template = VnfEm(**\{'vnf\_id':0, \\
	\> 'vnf\_cpu': 0.15, 'vnf\_ram': 100, \\
	\> 'vnf\_type': 'vnf\_matinv', \\
	\> 'vnf\_img': 'vnfserver:latest', \\
	\> 'vnf\_cmd': template\_cmd,'vnf\_port': 5000\}) \\
	nfvManager.add\_vnf\_into\_pool(vnfEm\_template) \\
	nfvManager.add\_vnf\_service\_into\_pool(0,1,\\
	\> **\{"band":0.5\})
\end{tabbing}

\subsubsection{Definition of SFC User Services}

Since the simulation mode requires a realistic data transmission process, it is necessary to define UE and add them to the manager pool of UEM. Mini-SFC also uses Docker containers to simulate user behavior, triggering the transmission of service traffic. Therefore, the template\_cmd here is similarly used as the command template for launching UE containers. The user equipment plays the role of initiating requests in the simulation. Here we also use an image prepared in advance. The simulator interacts with the server on the container to enable it to transmit a matrix to the VNF and initiate a calculation request.

\begin{tabbing}
	\hspace*{20pt}\= \kill
	\textbf{from} minisfc.mano.uem \textbf{import} UeManager, Ue \\
	ueManager = UeManager() \\
	template\_cmd = "python run\_command.py \\
	\> --ue\_name = \$ue\_name --ue\_type=\$ue\_type \\
	\> --ue\_ip = \$ue\_ip --ue\_port=\$ue\_port" \\
	ue\_template = Ue(**\{'ue\_id':0, \\
	\> 'ue\_type': 'ue\_post', \\
	\> 'ue\_img': 'ueserver:latest', \\
	\> 'ue\_cmd': template\_cmd, 'ue\_port': 8000\}) \\
	ueManager.add\_ue\_into\_pool(ue\_template) \\
	ueManager.add\_ue\_service\_into\_pool(0,1, \\
	\> **\{"req\_delay":1\})
\end{tabbing}

The step for selecting a solver is identical to it in Example A. Due to the solver's templated design, the same SFC orchestration algorithm can operate across different simulation modes with minimal or no modification. This simplifies the validation process for researchers developing their own algorithms.

\subsubsection{Integrating Components into the Engine to Launch the Simulation}

Create and initiate the Mini-SFC network simulation with the use\_container=True parameter configured to enable the container-based simulation functionality. This setting ensures that the simulation environment closely mimics the performance of an actual system, providing a more realistic testing ground.

\begin{tabbing}
	\hspace*{20pt}\= \kill
	\textbf{from} minisfc.net \textbf{import} Minisfc \\
	net = Minisfc(substrateTopo, serviceTopo, \\
	\> nfvManager, sfcSolver, ueManager, \\
	\> use\_container = True) \\
	net.start() \\
	net.stop() \\
\end{tabbing}

\begin{figure}[htbp]
	\centerline{\includegraphics[width=0.5\textwidth]{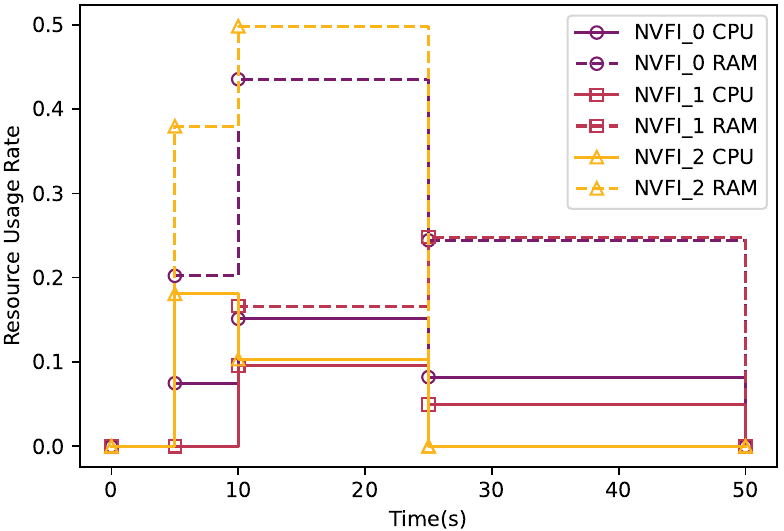}}
	\caption{The resource load status of NVFIs in the container-based simulation example.}
	\label{fig_4}
\end{figure}

In this example, we utilized the same experimental scenario and deployment strategy as in the numerical simulation to provide a more intuitive comparison with the results obtained from container-based simulation. During operation, real-time data on the network resource usage by VNFs simulated within various Docker containers was collected. This data was then mapped to their respective NVFI nodes and aggregated. The resulting resource load conditions monitored by each infrastructure manager are shown in Figure \ref{fig_4}.

It can be observed that node NVFI\_2 still has the highest RAM utilization rate as shown in Figure \ref{fig_4}. However, its CPU utilization rate becomes lower than that of NVFI\_0 after the 10th second. The primary reason for this difference is that, unlike in numerical simulations, container-based simulations can accurately reflect the congestion experienced by preceding VNFs deployed on NVFI\_0 due to substantial transmission traffic. Consequently, the computational workload of subsequent VNFs significantly decreases, reducing their CPU usage. Additionally, at the 25th and 50th seconds, as services conclude, the resources occupied by the containers are released, returning the network to its initial state.

In container-based simulation, Mini-SFC similarly supports hyperparameter exploration to adapt to specific use cases. Critical hyperparameters here include VNF CPU/RAM allocations, inter-VNF bandwidth limits, and user request intervals. For example, adjusting the vnf\_cpu parameter from 0.15 to 0.3 reveals how increased computational resources affect matrix inversion latency in the VNF, while modifying req\_delay demonstrates congestion impacts on service delivery. Such experiments are simplified by Mini-SFC’s unified interface: researchers only need to modify template parameters without altering core simulation logic, a contrast to vim-emu\cite{vim-emu}, where hyperparameter tuning requires deep integration with OSM workflows, increasing complexity.

The container-based simulation example highlights Mini-SFC's ability to mimic real deployment environments. By using Docker containers to simulate VNF lifecycle dynamics, Open vSwitch to manage network interactions, and user equipment containers to generate realistic service traffic, it accurately reflects how resource competition and network congestion impact SFC performance. In contrast, vim-emu\cite{vim-emu} supports containerized simulation and dynamic topology but imposes a steep learning curve due to its reliance on the complex OSM platform, requiring researchers to master intricate orchestration workflows. Apollo\cite{Apollo}, as a closed-source tool, lacks implementation details and cannot support custom algorithm validation. Mini-nfv\cite{Mini-NFV} fails to support dynamic topology, hindering its ability to simulate network structure changes. Mini-SFC’s advantage here is its balance of realism (via containerization) and usability (via modular design), while enabling seamless switching between numerical and container modes. This allows researchers to conduct end-to-end studies from theoretical design to practical deployment validation—flexibility unmatched by single-mode tools or overly complex platforms.

\section{Mini-SFC Application Cases}

We simulate a real event to validate our design is suitable for real-world deployment, with controlled scenario assumptionsto verify the framework’s basic effectiveness under dynamic topologies. The simulation scenario is inspired by the catastrophic ﬂoods that hit Henan Province of China in 2021, where communication infrastructures were severely destroyed. To alleviate the impact of the disaster, we assume a SAGIN isdeployed for disaster relief. Four orbits of the Kuiper constellation are set at an altitude of 590 km. 10 satellites are uniformly distributed in each orbit to form the satellite network layer. Additionally, ﬁve UAVs are placed on preplanned trajectoriesto serve as temporary communication nodes over the affectedareas, while three surviving ground base stations offer basic connectivity.

In terms of computational resources, space, air, and ground nodes are equipped with capacities of 3 Gb/s, 300 Mb/s, and 20 Gb/s, respectively. All nodes are furnished with 512Gb of memory. The bandwidths for space-ground and inter-satellite links are set as 200 Mbps and 500 Mbps, respectively. To capture dynamic changes in connectivity, the simulation operates for a duration of 10 hours (from 4:00 to 14:00), taking 60 snapshots of the network topology at 10 minute intervals. This conﬁguration ensures both robustness and adaptability, enabling SFCs tailored for URLLC, mMTC, and eMBB applications.

\begin{figure}[htbp]
	\centerline{\includegraphics[width=0.5\textwidth]{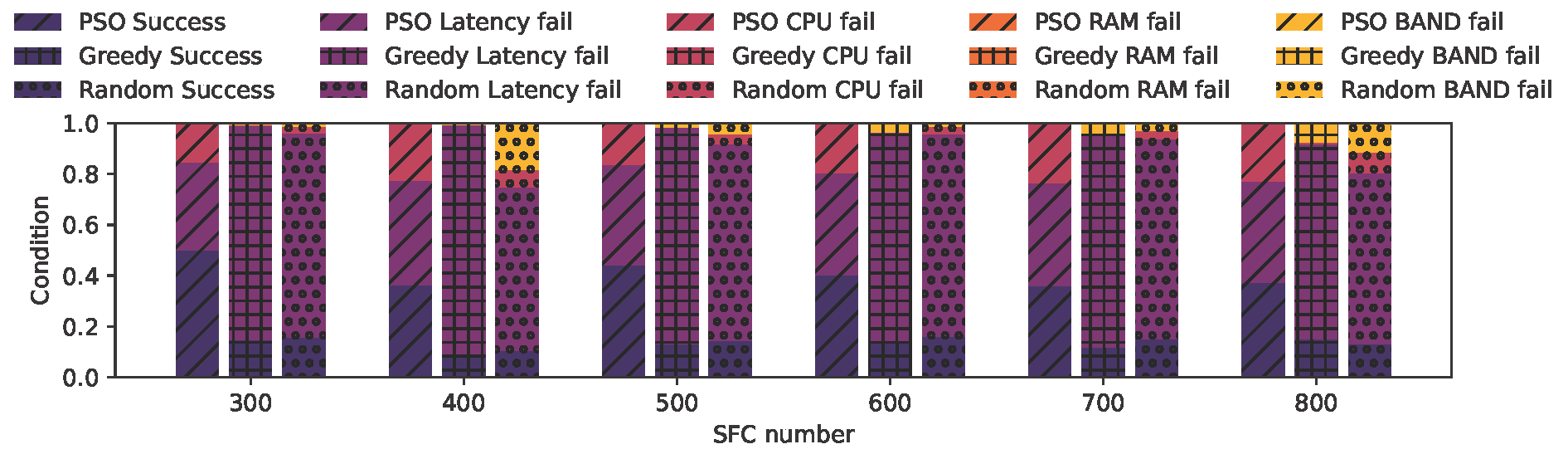}}
	\caption{Comparison of three algorithms for different numbers of SFCs in SAGIN scenario.}
	\label{fig_5}
\end{figure}

\begin{figure}[htbp]
	\centerline{\includegraphics[width=0.5\textwidth]{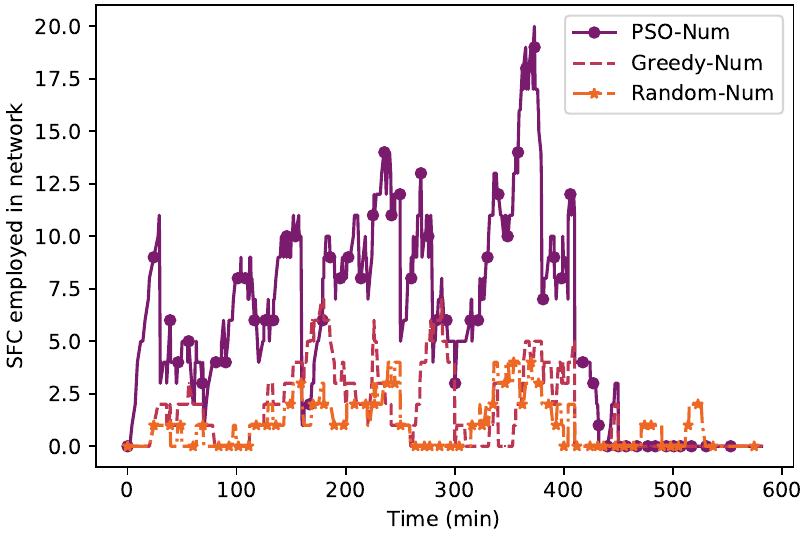}}
	\caption{The trend in the number of SFCs running during the simulation.}
	\label{fig_6}
\end{figure}

We implemented our previously published heuristic-based SFC orchestration algorithm\cite{PSO_SFC} to compare with other baseline algorithms within the Mini-SFC simulator. The outcomes are presented in Figures \ref{fig_5} and \ref{fig_6}. The Python script that implements all examples in this paper will not be presented due to page limitations but can be found in the reference\cite{Mini-SFC}.

Owing to the detailed logging capabilities of Mini-SFC, Figure \ref{fig_5} illustrates the performance of different algorithms under varying numbers of SFC requests and shows the specific reasons for deployment failures. Figure 6 shows the trend in the number of SFCs running during the simulation over time, whitch facilitates an easy comparison of the deployment algorithms' effectiveness in coping with network topology dynamics. With Mini-SFC, researchers can gain insights into how different algorithms manage SFC deployments under diverse conditions and topological shifts, thereby assessing their robustness and efficiency in dynamic environments.

The application of our heuristic algorithm\cite{PSO_SFC} in the SAGIN scenario is not intended to highlight algorithmic innovation, but rather to demonstrate Mini-SFC’s practical utility as a tool: it can seamlessly integrate and validate existing algorithms, log detailed performance metrics, and support comparisons under dynamic topology changes. This case study illustrates how Mini-SFC enables researchers to focus on algorithm design by handling the complexity of simulation setup, a capability absent in our earlier work which required custom simulation scripts.

\section{Conclusion and Future Work}

The modular simulation framework Mini-SFC provides a flexible experimental platform for the verification of orchestration and management strategies of SFCs, through the support of numerical and container-based virtual simulation. Numerical simulation plays a role in quickly verifying the performance of algorithms, while container-based virtual simulation is more conducive to reflecting complex interactions and resource contention issues. The design of Mini-SFC also takes into account the need for single-machine deployment, which is particularly beneficial for research teams with limited resources. By providing standardized algorithm templates and unified interfaces, it simplifies the integration of custom orchestration algorithms, allowing researchers to focus on innovation without being hindered by the complexity of the simulation platform.

The architecture of Mini-SFC ensures a high degree of scalability and potential integration with platforms such as Kubernetes (K8S), opening the way for enhanced physics simulation. In the future, we will further explore integration with K8S to enhance the scheduling capability of device clusters and provide a more realistic and reliable experimental environment for large-scale distributed systems. In addition, intelligent algorithms such as machine learning (ML) and deep learning (DL) are applied to optimize SFC management, which is expected to improve automation level and decision efficiency. These explorations are designed to enable researchers to better address the challenges of increasingly complex networks and lay a solid foundation for the close integration of cloud computing and communication in the future.

\section*{Acknowledgment}

This work has received support from the National Natural Science Foundation of China under Grant 62171158 and National Key Laboratory of Air-based Information Perception and Fusion and Aeronautical Science Foundation of China under Grant 20230001004002 and it was also supported by the Fundamental Research Funds for the Central Universities under Grant 2242022k60006.

\bstctlcite{IEEEexample:BSTcontrol}
\bibliographystyle{IEEEtran}
\bibliography{IEEEabrv,IEEEbib}

\end{document}